\documentstyle[12pt]{article}
\input{tcilatex}

\begin{document}

\title{Quantum $LC$ circuits with charge discreteness: normal and anomalous spectrum}
\author{J. C. Flores}
\date{Departamento de F\'{i}sica, Universidad de Tarapac\'{a}, Casilla 7-D, Arica,
Chile}
\maketitle

\bigskip

\bigskip

\baselineskip=16pt

\bigskip

{\bf Abstract}: A solvable quantum $LC$ circuit with charge discreteness is
studied. Two discrete spectral branches are obtained: (i) the normal branch
corresponding to a charged capacitor with integer effective charge $k=q_{e}n$
($q_{e}$ elementary charge, $n$ integer) and (ii) the anomalous branch where
the energy is related to non-integer effective charge $k=q_{e}(n-x)$ in the
capacitor. For usual mesoscopics data like quantum point contact we found $%
x\sim 0.7$.

\bigskip

\bigskip 
\[
\]

Published in Europhys.Lett. {\bf 69},116 (2005).



\newpage

In mesoscopic systems the discrete nature of the electrical charge plays an
important role [1]; phenomena like Coulomb blockage, current magnification,
spectral properties of wires etc. are related to it. In addition, the
concept of an elementary quantum flux \ $h/e$ arises naturally from charge
discreteness. On the other hand, some of the above mentioned systems, and
others, could be studied through the analogy with simple electrical circuits
with capacitances and inductances [2-7]. From this point of view, the
quantization of circuits with charge discreteness [3] plays an important
role in mesoscopic physics and this paper is related to this topic. We
emphasize that charge discreteness is assumed in this paper as an
experimental fact and therefore, it is not our goal to prove it from first
principles.

Consider a quantum $LC$ circuit with charge operator $\widehat{Q}$, magnetic
flux operator $\widehat{\phi }$ and the commutation rule $\left[ \widehat{Q},%
\widehat{\phi }\right] =i\hbar $. In close analogy with the known quantum
problem of a particle in a box, where the momentum is quantized, boundary
conditions on the state $\psi (\phi )$ must be imposed. \ So, on the wave
function we must impose the charge discreteness condition related to

\begin{equation}
\psi (\phi +\frac{h}{q_{e}})=\psi (\phi ),
\end{equation}
where $q_{e}$ is the elementary charge. Note that (1) means that the system
must be translation invariant in the $\phi $-space. Moreover, the
eigenvectors of the charge operator $\widehat{Q}=i\hbar \partial /\partial
\phi $ are $e^{-iq_{e}\phi j/\hbar }$, with eigenvalues $jq_{e}$ ($j$
integer ) and compatible with charge discreteness. In this way, the state of
the systems in charge representation $\psi _{j}$ and flux representation $%
\psi (\phi )$ are connected by the transformation

\begin{equation}
\psi (\phi )=\sum_{j}e^{-i\phi q_{e}j/\hbar }\psi _{j}.
\end{equation}

We note that, in the flux space, eqs. (1) \ and (2) are consequences of
charge discreteness and the fact that the charge operator and flux operator
are canonical conjugate.

The Hamiltonian $\widehat{H}$ for the $LC$ circuit with charge discreteness
requires compatibility with the boundary condition (1). Note that when $%
q_{e}=0$ the $LC$-Hamiltonian is given by

\begin{equation}
\widehat{H}_{q_{e}=0}=\frac{1}{2L}\widehat{\phi }^{2}+\frac{1}{2C}\widehat{Q}
^{2},
\end{equation}
where electron-electron interaction is automatically considered in the
capacitor properties.

The periodic boundary condition (1), tells us that the Hamiltonian has the
generic form 
\begin{equation}
\widehat{H}_{q_{e}}=\frac{2\hbar ^{2}}{q_{e}^{2}L}P(\frac{q_{e}}{h}\widehat{%
\phi })+\frac{1}{2C}\widehat{Q}^{2},
\end{equation}
where $P(x)$ is a periodic function of period $1$ (translation invariance).
The factor $2\hbar ^{2}/q_{e}^{2}L$ was incorporated to make contact with
references [3-7], were charge discretization was considered. In fact,
references [3-7] consider charge discreteness with the choice $P(x)=\sin
^{2}(\pi x/2)$. This is directly related to the discretization of the
operator $i\hbar \partial /\partial Q$ as finite differences. Nevertheless,
it is important to realize that there are other possible choices for
discretization procedure which could be used to represent charge
discreteness. Namely, any system described by the Hamiltonian (4), with the
additional condition (1) imposed upon it, represents a discrete charge
quantum $LC$ circuit. \ Note that the inductance $L$ and capacitance $C$ are
fixed parameters in our theory. Nevertheless, an equivalent approach could
be assumed if, for instance, the inductance in eq.(3) is flux-depending
(effective inductance) in correspondence with (4).

In this paper we shall study a solvable model for the $LC$ circuit,
described by a Hamiltonian with charge discreteness, and closely related to
quantum point contact systems. We shall consider a Hamiltonian with the
structure of (4) and related to the Schr\"{o}dinger equation in flux
representation ( $\widehat{Q}=-i\hbar \partial /\partial \phi $):

\begin{equation}
E\psi \left( \phi \right) =\frac{2\hbar ^{2}}{q_{e}^{2}L}\left\{
\sum_{l}\delta \left( \frac{q_{e}}{h}\phi -l\right) \right\} \psi \left(
\phi \right) -\frac{\hbar ^{2}}{2C}\frac{\partial ^{2}}{\partial \phi ^{2}}%
\psi \left( \phi \right) .
\end{equation}
The above equation is the basis of our calculations and is closely related
to the Kronig-Penney model used in Solid State Physics, nevertheless, the
condition (1) is more restrictive than the usual one (Bloch theorem). In
fact, translation invariance of (5) tell us that $\psi (\phi +h/{q_{e}}%
)=\psi (\phi )exp(iK2\pi /q_{e})$ (Bloch) and then from (1) we must consider
only $K=0$. \ \ So, some points, or levels, of the band structure of (5) are
related to the condition of charge discreteness.

As usual for an equation like (5), between two $\delta -$barriers (at
position $l/hq_{e}$ and $\left( l+1\right) /hq_{e}$, $l$ integer) we have
the equation $E\psi \left( \phi \right) =-\frac{\hbar ^{2}}{2C}\frac{%
\partial ^{2}}{\partial \phi ^{2}}\psi \left( \phi \right) $ \ and its
solution is given by

\begin{equation}
\psi _{l}(\phi )=A_{l}e^{ik\phi /\hbar }+B_{l}e^{-ik\phi /\hbar },
\end{equation}
where the {\it effective charge} $k$ is a parameter with dimension of
electrical charge. In term of this parameter, the energy of the system
becomes

\begin{equation}
E=\frac{1}{2C}k^{2},
\end{equation}
corresponding to the energy of a capacitor. As a consequence of the boundary
condition (1), the parameter $k$ is not continuous. \ In fact, the matching
conditions in a barrier\ $\psi (\frac{h}{q_{e}}l)_{+}=\psi (\frac{h}{q_{e}}%
l)_{-}$ (by left and right side) and the step in the first derivative \ \ $%
\psi ^{\prime }(\frac{h}{q_{e}}l)_{+}-\psi ^{\prime }(\frac{h}{q_{e}}%
l)_{-}=\left( 4Ch/Lq_{e}^{3}\right) \psi $ \ \ tell us that \ the
coefficients $A$ and $\ B$ of (6) are not arbitrary. For instance, for the
coefficient $A$ one obtains the recursive equation

\begin{equation}
A_{l+1}e^{2\pi i\frac{k}{q_{e}}}+A_{l-1}e^{-2\pi i\frac{k}{q_{e}}%
}=A_{l}\left( \alpha e^{2\pi i\frac{k}{q_{e}}}+\alpha ^{\ast }e^{-2\pi i%
\frac{k}{q_{e}}}\right) ,
\end{equation}
where the coefficient $\alpha =1-i\frac{2C\hbar h}{Lq_{e}^{3}k}$. Defining
the new variable $x_{l}=e^{i2\pi lk/q_{e}}A_{l}$ the above equation becomes

\begin{equation}
x_{l+1}+x_{l-1}=\left\{ 2\cos \left( \frac{2\pi k}{q_{e}}\right) +\frac{%
4C\hbar h}{Lq_{e}^{3}k}\sin \left( \frac{2\pi k}{q_{e}}\right) \right\}
x_{l}.
\end{equation}
Like to the Solid State case, equation \ (9) gives a band structure when the
condition

\begin{equation}
\cos \left( \frac{2\pi }{q_{e}}K\right) =\cos \left( \frac{2\pi k}{q_{e}}%
\right) +\frac{2C\hbar h}{Lq_{e}^{3}k}\sin \left( \frac{2\pi k}{q_{e}}%
\right) ,
\end{equation}
holds for some values of $K$. Since the energy $E$ and the parameter $k$ are
related by (7), the above equation gives formally the spectrum $E(K)$ of
(5). Nevertheless, as said before, from the condition (1) for quantum
circuits we must consider $K=0$ \ on (10). Therefore, the spectrum of the $%
LC $ circuit with charge discreteness is obtained from the solutions of

\begin{equation}
1=\cos \left( \frac{2\pi k}{q_{e}}\right) +\frac{2C\hbar h}{Lq_{e}^{3}k}\sin
\left( \frac{2\pi k}{q_{e}}\right) .
\end{equation}
Observe that $k=0$ is not a solution (zero point fluctuations). To solve
(11), consider the definition

\begin{equation}
\tan \theta =\frac{2C\hbar h}{Lq_{e}^{3}k},
\end{equation}
then, the equation (11) becomes

\begin{equation}
\cos \left( \theta -\frac{2\pi k}{q_{e}}\right) =\cos \theta .
\end{equation}
There is two solutions:

{\bf (a)} {\bf Normal spectrum (}$E^{(N)}${\bf ):} where $\left( \theta
-2\pi k/q_{e}\right) =+\left( \theta -2\pi n\right) $, then we have the
effective charge $k=nq_{e}$ $(n\neq 0,$ and integer$)$ and the spectrum is

\begin{equation}
E_{n}^{(N)}=\frac{1}{2C}\left( nq_{e}\right) ^{2},
\end{equation}
with some similarities to the spectrum of a quantum dot [1].

{\bf (b) Anomalous spectrum \ (}$E^{(A)}${\bf )}: where $\ \ \left( \theta
-2\pi k/q_{e}\right) =-\left( \theta -2\pi n\right) $, then $\theta =\pi
\left( k/q_{e}-n\right) $. Using the definition (12), in this case the
equation for the effective charge $k$ becomes

\begin{equation}
\frac{k}{q_{e}}\tan \left( \frac{k}{q_{e}}\pi \right) =\frac{2C\hbar h}{%
Lq_{e}^{4}},
\end{equation}
which defines implicitly the anomalous spectrum $E_{n}^{(A)}$. This is an
interesting result, because our technique produces a second branch different
from the expected result (14) breaking degeneracy. For instance, in the
approximation $2C\hbar h/Lq_{e}^{4}\gg 1$ \ we obtain for the effective
charge $k\sim q_{e}\left( n-1/2\right) $, with its corresponding energy
spectrum given by (7). \ In \ the anomalous case we have $0<k/q_{e}<0.5$
(mod.1). In general, from the above results we conclude that between two
integer (normal) solutions of (11) (given by $k=nq_{e}$) \ there is another,
(anomalous) related to the branch (15). \ In this way, the magnetic term in
equation (5) \ removes degeneracy and one obtains the two branches \ at the
spectrum. Moreover, for large values of $k$ \ \ this two branches approach
together as shows figure 1.

Equation (15) \ can be simplified by assuming that the effective charge $%
k=q_{e}(n-x)$ where the approximation $n\gg x$ is considered ( $0<x<1$). In
this approximation, equation (15) becomes

\begin{equation}
n\tan \pi x=-\frac{2C\hbar h}{Lq_{e}^{4}}.
\end{equation}

As a possible application of the above results consider a two-dimensional
electron gas in a mesoscopic device like a quantum point contact [8-13].
Since crudely $C\sim \epsilon _{o}d$ and $L\sim \mu _{o}d$ then the right
hand side of (15) is $2C\hbar h/Lq_{e}^{4}\sim 1,399\times 10^{3}$[14]. On
the other hand, this two-dimensional gas have a density of $10^{15}m^{-2}$
[8], then the number $n$ of particles can be estimated as $10^{15}\times
\left( 10^{-6}\right) ^{2}=10^{3}$(with $d\sim 10^{-6}m,$ a micrometric
distance [8]). The explicit evaluation of (16) is

\begin{equation}
10^{3}\tan \pi x=-1.399\times 10^{3},
\end{equation}
with the solution $x=0.698\sim 0.7$. Therefore, just below the normal state
with effective charge $k=q_{e}n$, there lies an anomalous state with
effective charge $k=q_{e}(n-x)$. \ Note that strictly the right hand of \
eqs. (15) or (16) are depending on the geometry \ (size, etc.) of the
systems and the assumptions \ $C\sim \epsilon _{o}d$ and $L\sim \mu _{o}d$
is a crude approximation. Namely, strictly we can expect a size dependence
on $x$.

Decoherence and dissipation is a hard topic in quantum mechanics and then,
also in quantum circuit. In fact, circuits are close related with effects
like Ohm law or Joule dissipative effects. The standard way to consider
decoherence and dissipation in \ quantum mechanics is related to the
connection of the systems to a thermic bath. Closed equations \ for the
system are found by tracing onto the bath degree. This standard technique
was used in meso-particles ([15] and [16-20]). In our case of circuits, \ a
similar procedure could be envisaged in the future.

In resume, \ we have presented a quantum solvable $LC$ circuit with charge
discreteness. The spectrum was studied and it contains two branches, the
normal \ (14) and the anomalous (15). It corresponds \ to the remotion of
degeneracy due to the magnetic term in (5). For quantum point contact data,
the spectrum was \ explicitly characterized (17). From a general point of
view, our theory \ represents an easily and efficient approach to consider
mesoscopic and nano-devices in the future.

\bigskip

This work was supported by FONDECYT\ (Grant \ 1040311). Useful discussions
with C. A. Utreras (UACH), A. Perez (PUC) and S. Montecinos (UFRO)
concerning charge discreteness and boundary conditions are acknowledged.

\newpage 

\end{document}